\documentclass[12pt,preprint]{aastex}

\shorttitle{The Cornell High-order Adaptive Optics Survey for Brown
  Dwarfs in Stellar Systems--II:  Results from Monte Carlo Population Analyses}
\shortauthors{Carson et al.}

\begin{document}

\title{The Cornell High-order Adaptive Optics Survey for Brown Dwarfs
  in Stellar Systems--II:  Results from Monte Carlo Population Analyses}

\author{J. C. Carson\altaffilmark{1,2}, S. S. Eikenberry\altaffilmark{3}}
\affil{Department of Astronomy, Cornell University, Ithaca, NY 14853}

\author{J. J. Smith\altaffilmark{4}}
\affil{Jet Propulsion Laboratory, California Institute of Technology, 4800 Oak Grove Drive, Pasadena, CA 91109}

\and

\author{J. M. Cordes}
\affil{Department of Astronomy, Cornell University, Ithaca, NY 14853}

\altaffiltext{1}{present address: Jet Propulsion Laboratory, Earth \&
  Space Sciences, 4800 Oak Grove Dr., MS 183-900, Pasadena, CA 91109}
\altaffiltext{2}{Postdoctoral Scholar, California Institute of Technology.}
\altaffiltext{3}{present address: Department of Astronomy, University of Florida, 211 Bryant
  Space Science Center, Gainesville, FL 32611}
\altaffiltext{4}{present address: University of California, Berkeley, CA 94720}

\begin{abstract}
     In this second of a two-paper sequence, we present Monte Carlo population simulation results of brown dwarf companion data collected during the Cornell High-order Adaptive Optics Survey for brown dwarf companions (CHAOS).  Making reasonable assumptions of orbital parameters (random inclination, random eccentricity and random longitude of pericentre) and age distributions, and using published mass functions, we find that the brown dwarf companion fraction around main sequence stars is 0.0\%-9.3\% for the 25-100 AU semi-major axis region.  We find a corresponding L-dwarf companion fraction of 0.0\%-3.3\%.  We compare our population analysis methods and results with techniques and results presented by several other groups.  In this comparison we discover that systematic errors (most notably resulting from orbital projection effects) occur in the majority of previously published brown dwarf companion population estimates, leading authors to claim results not supported by the observational data.
\end{abstract}


\section{Introduction}

 A deeper understanding of the formation and evolution of stars and
 planets is a leading goal of modern astronomy.  The
 understanding of brown dwarfs, objects that make up the transition
 mass between stars and planets, is a necessary step in achieving this
 goal.  The brown dwarf category typically divides into two subgroups, L-type and T-type:  T-dwarfs make up the cooler subclass, usually defined by the existence of methane in their atmospheres;  L-dwarfs are usually defined by atmospheres cool enough to possess lithium, unlike most warmer M-stars, but warm enough that the vast majority of their atmospheric carbon is locked in CO rather than CH$_{4}$.  While diverse theories abound
 (Reipurth \& Clarke [2001], Bate, Bonnell \& Bromm [2002], Boss
 [2001], for example) to explain the formation of such objects,
 accurate brown dwarf population statistics that may test these
 predictions remain scarce.  To help address this gap, we initiated CHAOS, the Cornell High-order Adaptive
     Optics Survey for brown dwarf companions.  In this survey, we used the Palomar Hale
     Telescope's PALAO Adaptive Optics System (Troy et al. 2000) and
     accompanying PHARO science camera (Hayward et al. 2001) to
     conduct high-contrast coronagraphic observations of 80
     main-sequence stars out to 22 parsecs (Carson [2005], Carson
     et al. [2005]).  The most prominent result
     from this investigation was that zero systems showed positive
     evidence for brown dwarf companions, a result consistent with the
     "brown dwarf desert" described by Marcy \& Butler (2000) for
     narrow separations ($<$3 AU).  But while our results are
     consistent with a brown dwarf desert, do they necessarily assert
     that a brown dwarf desert exists at intermediate separations (like
     25-100 AU)?  Clearly, there are a number of circumstances that
     might entail an intermediate separation brown dwarf existing
     around a target star, while our observations of that star lead to
     a null detection:  the brown dwarf's brightness could be below
     our sensitivity threshold;  the brown dwarf might be projected behind the parent star;  it might have an exceptionally high eccentricity causing its typical separation to differ significantly from its semi-major axis.  These and other factors lead to a non-trivial relationship between observed and physical companion fractions.  In this paper, we investigate this relationship in order to determine the brown dwarf companion population range consistent with our observational results.

The Monte Carlo simulations described in this paper employ the following strategy: 1) gather all the simulated orbits for a given range of eccentricity, inclination, longitude of pericentre, and semi-major axis and project them onto an individual star field of view; use this exercise to calculate the odds that the orbiting object falls within the field of view; 2) making use of target star noise maps (generated from observational data), determine the odds that a given apparent K-magnitude orbiting object would be detectable, assuming it lies within the field of view; 3) convert the apparent K-magnitude to a brown dwarf mass, using the known target star distance, a sample age value, and Baraffe et al. (2003) brown dwarf evolutionary models; 4) assuming a sample brown dwarf mass function and a random inclination, eccentricity, and longitude of pericentre, determine the net likelihood, for a given semi-major axis, that an orbiting brown dwarf (assumed to be existent) would be detected during the target star observation; 5) repeat this procedure for all target stars to find a net survey detection probability; 6) use a likelihood analysis approach to determine which brown dwarf companion fractions are consistent with the derived detection probability and our null result.
  
In the literature of brown dwarf companions to main sequence stars, we are aware of only four other published studies that attempt to derive rigorous population statistics: Marcy \& Butler (2000), McCarthy \& Zuckerman (2004), Gizis et al. (2001), and Lowrance (2005).  Out of these four studies, only one, Marcy \& Butler (2000), considers non-zero inclinations and eccentricities in their models.  Our Monte Carlo simulations, providing a thorough analysis of non-zero inclinations and eccentricities, therefore provide an important contribution to our knowledge of brown dwarf companion frequencies, and brown dwarf formation as a whole.  

   Section 2 describes our target sample.  Section 3 summarizes our
   simulation procedures.  Sections 4 and 5 present our results and conclusions. 

\section{Target Sample}
   In the first paper of this series (Carson et al. 2005), we listed
   individual information on all target stars in our sample.  There,
   from our 80 star sample, we compiled 3 A stars, 8 F stars, 13 G
   stars, 29 K stars, 25 M stars, and 2 stars with ambiguous spectral
   types.  Six of these targets were published binaries.  The \textit{effective} spectral type coverage, however, depends on the relative sensitivities of the various target observations.  We determine the \textit{effective} spectral type coverage by first calculating, from our observational data, the median sensitivity for a 13-73 M$Jup$ object orbiting with a semi-major axis 25-100 AU, for all targets with a given spectral type.  (See Section 4.1 for a detailed presentation of our sensitivity calculations.)  We take the ratio of this coverage (e.g. 40\% coverage, 80\% coverage, etcetera) to the median survey coverage and multiply it by the number of stars with the given spectral type.  Doing so, we derive an \textit{effective} target distribution of 1.9 A stars, 5.9 F stars, 16.1 G stars, 28.6 K stars, 25.6 M stars, and 1.9 stars with ambiguous spectral types.  

To gain some information on the age range of
   this sample, we examine Fe/H line strengths published in Cayrel de
   Strobel et al. (1997) and Cayrel de Strobel, Soubiran, \& Ralite
   (2001).  Table 1 displays Fe/H line
   strengths for the quarter of stars in our sample that had published
   values.  To derive ages from these line strengths, we use the following equation from Rocha-Pinto et al. (2000):

\begin{equation}
         t = 10(0.44 - [Fe/H])
\end{equation}

\textit{t} is the age in Gyrs.  We assume a minimum age of 250 Myr and
a maximum age of 10 Gyr.  For the stars with published line
strengths, such an equation yields an age distribution of 4.8 $\pm$
2.7 Gyr.  We note here that Equation 1 represents a loose constraint on
stellar age.  However, for the purposes of our simulations, it
represents a useful approximation.  To extrapolate these values to the
complete target sample, we must ignore observational biases in the 
Fe/H line strength table.  For instance, fainter stars such as
M-dwarfs are under-sampled, considering their frequency, while
brighter stars are somewhat over-sampled.  Stellar age, however,
should not correlate strongly with spectral type, outside of the rarer
early-type stars.  Thus, for the purposes of our simulations, we
extrapolate our Table 1 ages to be indicative of the complete
sample.  To address potential errors in this assumption, we include an
examination describing how
errors in these age estimates might affect our outputted values (see Section
4.2).    

  Prominent alternative techniques for determining stellar ages take advantage, for instance, of Ca II emission, lithium abundance, X-ray activity, galactic space motion, and galactic birth models.  Ca II emission, lithium abundance, and X-ray activity are useful for young stars (less than a couple hundred Myrs typically), but provide poor constraints for older target populations (see Song, Zuckerman, \& Bessel [2004] for a more thorough discussion of these age markers).  Galactic space motion can be effective for both old and young stars, but given that most, if not all, of our target population have poorly defined motion associations, this technique is impractical for our purposes.  Galactic birth models provide age predictions without the aforementioned drawbacks.  However, given the degree of controversy accompanying the competing birth models, we conclude that their predictions are not better than the loose constraints provided by iron line estimates.  (See Burgasser [2004] for a more thorough discussion of galactic birth rate models as an estimate of stellar ages.)       

     As an additional caveat, we note that there is no a priori reason
     to expect that the brown dwarf companion frequency presented for
     this largely single-star sample
     is equivalent to the brown dwarf companion frequency around
     double, triple, or quadruple systems.  Indeed, some brown
     dwarf formation scenarios, such as Clarke, Reipurth, \&
     Delgado-Donate (2004),
     predict that the brown dwarf companion frequency around multiple
     systems will differ significantly from the single star companion
     frequency.  Estimating a multiple system brown dwarf companion
     frequency is therefore beyond the scope of this paper.

\section{Simulation Procedures}

\subsection{The Mathematics of Our Likelihood Analysis Approach}
     Our Likelihood Analysis approach seeks to answer the question: How consistent is a hypothetical brown dwarf companion population with our observed results?  To answer this question, we begin by defining two parameters, \textit{$\theta$} and \textit{$\phi$}, which represent vectors describing all of the parameters of our brown dwarf population.  \textit{$\phi$} describes the fraction of stars with brown dwarf companions.  \textit{$\theta$} represents all other population parameters including brown dwarf masses, brown dwarf luminosities, orbital radius distributions, etcetera.  Next we note that, for each star, we have a measurable probability of detecting a given luminosity brown dwarf.  This probability depends on \textit{D}, distance to the system, \textit{L}$_{*}$, luminosity of the parent star, \textit{r}, projected orbital separation, and \textit{I}, instrumental parameters.  \textit{I} includes factors such as throughput, exposure time, PSF stability, etcetera.  For a given star, therefore, we may write the probability of a brown dwarf detection as:

\begin{equation}
         P_{detection} = P(\phi, \theta, D, L_{*}, I)
\end{equation}     

Since we detected no brown dwarf companions in our survey, we may write that, for the \textit{j}th observed star, the number of expected detections is \textit{$\mathcal{N}_{j}$} = \textit{P$_{detection-j}$} $\ll$ 1.  Poisson statistics then dictate that the probability of detecting \textit{k} brown dwarfs around a given star is:

\begin{equation}
         \mathcal{P}_{k} = \frac{\mathcal{N}^{k}}{k!} e^{-\mathcal{N}}
\end{equation}  

Since we detected zero brown dwarfs, we seek to determine a likelihood for \textit{k} = 0.  Taking the product of \textit{P$_{detection}$} over all 80 sampled stars, with \textit{k}=0, we may derive the following likelihood equation: 

\begin{equation}
         \mathcal{L}(\phi,\theta,D,L_{*},I) = \prod_{j=1}^{80} e^{-\mathcal{N}_j}
\end{equation}

Recall that \textit{$\phi$} = fraction of stars with brown dwarf companions;  \textit{$\theta$} = all other brown dwarf population parameters (mass, luminosity, orbital radius, etcetera);  \textit{D} = distance to target system;  \textit{L$_{*}$} = luminosity of the parent star;  \textit{I} includes instrument parameters such as exposure time and throughput;  \textit{$\mathcal{N}_{j}$} = number of expected detections around the \textit{j}th star.  Our relation here therefore represents the likelihood of making zero detections through 80 observations of a brown dwarf companion population represented by \textit{$\theta$} and \textit{$\phi$}.  \textit{D}, \textit{L$_{*}$}, and \textit{I} are all measurable values that we define for each observing set.  Hence, they are factors included in \textit{$\mathcal{N}_j$}.  For the purposes of a straightforward analysis, we shall set \textit{$\theta$} to sample test populations.  We shall then let \textit{m$_j$} equal the segment of \textit{$\mathcal{N}_j$} which includes all parameters except for \textit{$\phi$}.  In other words, \textit{$\mathcal{N}_{j}$} = \textit{$\phi$m$_j$}.  Substituting this relation into equation 3 we now have  

\begin{equation}
         \mathcal{L}(\phi) = \prod_{j=1}^{80} e^{-{\phi}m_j}
\end{equation}

We note here that this analysis assumes that each star has a maximum of one orbiting brown dwarf.  A star with two brown dwarf companions would therefore be treated, statistically, as there being two stars in the sample which each possessed a brown dwarf companion.  This type of treatment takes precedent in population statistical studies such as Tokovinin (1992).  We recommend that the reader refer to that publication for a further discussion of such a technique. 

     For our final analysis, we choose a 90$\%$ confidence level for our conclusions.  Therefore, we seek to find the range of \textit{$\phi$} values which yield a null detection (through 80 observations) 90$\%$ of the time.  Clearly, \textit{$\phi$} = 0 will be the lower limit for \textit{$\phi$}.  $\phi_{u}$ represents the upper limit to the number of stars with brown dwarf companions.  Combining these facts with equation 4, we may derive the relation:  

\begin{equation}
         0.9 = \frac{\int\limits_0^{\phi_{u}}e^{-{\phi}({m_{1}+m_{2}+{\cdot}{\cdot}{\cdot}+m_{80}})}d\phi}{\int\limits_0^{\infty}e^{-{\phi}({m_{1}+m_{2}+{\cdot}{\cdot}{\cdot}+m_{80}})}d\phi}
\end{equation}   

Solving for \textit{$\phi_{u}$}, we get the simple equation:

\begin{equation}
         \phi_{u} = \frac{-ln(0.1)}{m_1+m_2+{\cdot}{\cdot}{\cdot}+m_{80}}
\end{equation}

In the next section, we describe how we may use Monte Carlo simulations to evaluate the \textit{m$_j$} values in equation 7.

\subsection[Sampled Brown Dwarf Orbits]{Sampled Brown Dwarf Orbits}
      As mentioned in Section 1, our Monte Carlo simulations begin by selecting, from a database of simulated orbits, all cases that cover a given range of semi-major axis, eccentricity, inclination, and longitude of pericentre.  While our database does not cover an infinite range of orbits, it does include all combinations of eccentricity from 0 to 0.9 (at 0.1 intervals), longitude of pericentre from 0 to 350 degrees (at 10 degree intervals), inclination from 0 to 90 degrees (at 10 degree intervals) and semi-major axis from 5 to 900 AU (at log[AU] intervals $\sim$0.03).  Each simulated orbit array records an orbital path as well as the fractional time that an orbiting object would spend at each path position (i.e. periastron positions have shorter fractions, apastron positions  have longer fractions).  In the next section we describe how we integrate these orbital arrays to derive survey detection probabilities.  

\subsection[Convolving Simulated Orbits With Target Fields of View and Observational Noise Maps]{Convolving Simulated Orbits With Target Fields of View and Observational Noise Maps}

     After gathering a set of potential orbits, we project our orbital arrays onto a target star field of view and determine the odds that an orbiting object with that semi-major axis would reside within the detector field of view.  Next we consider the likelihood, for instances when the object resides in the field of view, that it resides within a given noise threshold level, defined as a function of detector position, by our observational noise maps. (Note that the noise threshold varies strongly with detector position;  at the parent star center, it is effectively infinite; toward the edges of the field of view, it is typically flux-limited;  see Paper 1, Section 4.2 for a further discussion of observational noise maps).  This exercise therefore lets us determine the relative fractions of its period that a hypothetical brown dwarf spends at a given noise threshold position.  To determine the resultant detection probability, we select potential brown dwarf absolute K$_{s}$ magnitudes, extending from 8 to 23 magnitudes at intervals of 0.3 magnitudes.  (Brown dwarfs with absolute K$_{s}$ magnitudes brighter than 8 magnitudes are deemed, statistically, to be effectively non-existent;  brown dwarfs with absolute K$_{s}$ magnitudes fainter than 23 magnitudes are assumed to be invisible to our survey.)  We convert these absolute magnitudes to detector counts using the procedure described in Paper 1, Section 4.2.  We then determine, for each absolute K$_{s}$-magnitude, semi-major axis, and target observation, the odds of our being able to detect (at 5$\sigma$) an orbiting brown dwarf, assumed to be existent.  To determine the net survey detection probability, for each absolute K$_{s}$-magnitude and semi-major axis, we take the median, for a given K$_{s}$-magnitude and semi-major axis, of all the aforementioned odds values.   

\subsection[Convolving Detection Probabilities With Brown Dwarf Luminosity Functions]{Convolving Detection Probabilities With Brown Dwarf Luminosity Functions}

A goal of our analysis is to predict the overall detection probability for a brown dwarf, at a given semi-major axis, with a mass representative of the general 13-73 M$_{Jup}$ brown dwarf population. To do this, we need to conduct a weighted average of the probabilities associated with the different absolute K$_{s}$ magnitudes.  We do this by using a field brown dwarf mass function from the literature (see Section 4.1 for a discussion of the different mass functions we used).  We convert this mass function to a luminosity function using Baraffe et al. (2003) evolutionary models and a Table 1 age distribution.  (We start with published mass functions, rather than published luminosity functions, since brown dwarf mass functions should be roughly age independent.)  Using our resultant luminosity function, we are able to  conduct a weighted average over all sampled absolute K$_{s}$ magnitudes.  Thus, our final value represents a detection probability for a typical 13-73 M$_{Jup}$ orbiting brown dwarf at a given semi-major axis.

To determine a detection probability for the L-dwarf subset, we use luminosity functions published in Cruz et al. (2003).  Since the L-dwarf subset is defined by luminosity, we are able to avoid the age and mass function estimates used in the previous calculation.  Weighting the different K$_{s}$-magnitude probabilities according to the luminosity function, we determine a survey L-dwarf detection probability for each sampled semi-major axis.  

\subsection[Converting Detection Probabilities to Companion Fractions]{Converting Detection Probabilities to Companion Fractions}

Recall in Section 3 that we defined a value \textit{m$_j$} to be the probability of detecting an orbiting brown dwarf (assumed to be existent) around a given target.  The detection probability we calculated above is effectively an \textit{average} \textit{m$_j$}, or \textit{m$_{avg}$}, for a given semi-major axis.  We may therefore substitute \textit{m$_{1}$ + m$_{2}$ + $\cdot$ $\cdot$ $\cdot$ + m$_{80}$}, in equation 7, with 80\textit{m$_{avg}$} to determine \textit{$\phi$$_{u}$}, the companion fraction upper limit. 

\section{Population Simulation Results}

     In the following discussion of the population simulation results
     we address three questions:  (1) What are the sensitivity levels
     of our survey with regards to physical population characteristics
     (i.e. absolute magnitudes, semi-major axes, brown dwarf masses,
     etcetera)?    (2) What population upper limits are consistent
     with the observational data?  (3) How do our population analysis
     techniques and results compare with other brown dwarf companion
     population studies?  With regards to the first question, our
     simulations reveal that our survey is most sensitive to brown
     dwarfs orbiting with semi-major axes ranging from 25-100 AU.
     Given our sensitivity limits, our survey should detect 32\% of
     all 13-73 Jupiter mass objects in this semi-major axis regime,
     assuming Baraffe et al. (2003) evolutionary models, an age
     distribution as described in Section 2, and a brown dwarf mass
     function compiled from Slesnick, Hillenbrand, \& Carpenter (2004), Luhman (2004), Luhman et al. (2003), Luhman et al. (2005), and Muench et al. (2002) data.  From our population upper limit analysis we derive that, with a
     90\% confidence level,
 {\bf at most 9.3\% of our target stars possess brown dwarf companions with semi-major axes between 25 and 100 AU}.  If we choose a 95\% confidence level, our upper limit is 12.1\%.  To place these numbers in context, we note that Duquennoy \& Mayor (1991) and Fischer \& Marcy (1992) data indicate that just over 10\% of main sequence stars have 0.08-0.32 M$_{\odot}$ companions between 25 and 100 AU.  

If we only consider the L-dwarf subset, we determine a survey detection sensitivity of 86\% for the 25-100 AU semi-major axis region.  Accordingly, we determine an L-dwarf companion upper limit of 3.3\%.  (For a 95\% confidence level, the upper limit is 4.4\%.)  The following paragraphs describe our analysis techniques in greater detail as well as a comparison with other published surveys.

\subsection[Survey Sensitivities]{Survey Sensitivities}
     In Paper 1, figures 3 and 4, we described our survey sensitivities in
 terms of observational characteristics such as apparent magnitude and
 angular separation from the parent star.  
 But while such a description is revealing, it does not show sensitivities in
 terms of physical characteristics such as brown dwarf mass and
 semi-major axis.  In order to determine such relations we require
 output from Monte Carlo simulations.

Our baseline Monte Carlo parameter set includes a random inclination\footnote{Note that random inclination does not imply that face-on orbits are as common as edge-on orbits.  Whether an orbit appears face-on or edge-on depends also on orbital projection effects, which our simulations take into account.}, random eccentricity, random longitude of pericentre, Baraffe et al. (2003) evolutionary models, an age distribution derived from Table 1 values, and a brown dwarf mass function compiled from the median-combination of Orion Nebula Cluster (ONC; Slesnick et al. 2004), Taurus (Luhman, 2004), IC 348 (Luhman et al. 2003), Chamaeleon I (Luhman et al. 2005), and Trapezium (Muench et al. 2002) data.  Using this parameter set, we derive the thick solid detection probability curve plotted in Figure 1.

To determine the effect of an error in our assumed median age, we determine detection probabilities when our median age skews by $\pm$2 Gyr.  The \textit{dot-dashed} curves in Figure 1 display the results of such an examination.

Our use of mass functions derived from ONC, Taurus, IC 348, Chamaeleon I, and Trapezium clusters makes a non-ideal approximation;  such brown dwarf mass functions could differ from a brown dwarf mass function derived from solar neighborhood field brown dwarfs.  The disadvantage, however, of using a published solar neighborhood mass function (like Burgasser, 2004, for instance) is that there are typically higher uncertainties for the lower mass region.  The aforementioned clusters, in contrast, possess reasonable coverage across the entire brown dwarf mass regime, since they sample younger, brighter target populations.  Hence, despite their drawbacks, we prefer them to the higher uncertainty solar neighborhood mass function data.

An inspection of the published literature shows that the five cluster mass functions are not completely identical.  For our baseline simulation, we used a median-combination of the five mass functions.  But to show the impact of mass function differences on derived probability curves, we also plot detection odds using each of the individual mass functions (see \textit{dashed} curves in Figure 1). 

We also consider the effects of a strongly skewed eccentricity distribution.
\textit{dot-dot-dot-dashed} curves in Figure 1 display the effects on detection probabilities if we set eccentricity to either 0 or 0.9, instead of a random distribution.

We also determine sensitivities for the brighter L-dwarf subset of
brown dwarfs.  Since L-dwarfs are categorized by brightness and
composition, rather than mass or age, we do not require theoretical
evolutionary models, age estimations, or published mass functions for this analysis.  Figure 2
displays resulting detection probability curves, using Cruz et al. (2003) luminosity functions.  The \textit{dashed} curves represent results when we set the eccentricity to either 0 or 0.9.

\subsection[Population Upper Limits]{Population Upper Limits}\label{Population Upper Limits}

     To arrive at brown dwarf companion population upper limits, we
     combine the sensitivity data from the previous section with
     the upper limit calculation described in Equation 7.  As we mentioned in sections 3.1 and 3.5, \textit{m$_{j}$} represents the
     detection odds for a given star and a given brown dwarf semi-major
     axis.  Since Figures 1 and 2 display the consolidated, or mean
     detection odds for the whole survey, we may substitute ``{\textit
     m$_{1}$+m$_{2}$+$\cdot$ $\cdot$ $\cdot$+m$_{80}$}'' with
     ``80{\textit m$_{mean}$}'' (where ``80{\textit
     m$_{mean}$}'' is represented by the y-axis values in Figure 1 or 2) to
     determine the net brown dwarf companion fraction upper limits.  Figure 3 displays the results of such an analysis.
     From the plot we see that our survey is most sensitive to
     semi-major axes ranging from about 25 to 100 AU;  the plot indicates that, at a 90\% confidence
     level, at most 9.3\% of CHAOS targets have a brown dwarf
     companion between 25 and 100 AU.  (This value becomes 12.1\% for a 95\% confidence level.)   As a comparison, just over 10\% of main sequence stars have a 0.08-0.32 M$_{\odot}$ companion between 25 and 100 AU (according to Duquennoy \& Mayor 1991 and Fischer \& Marcy 1992 data).  Our upper limit results for alternate input parameters
     (i.e. altering the median age, changing the eccentricity
     distribution, etcetera) are described by the over-plotted curves.  

We also compute the brown dwarf companion fraction upper limit for the
L-dwarf subset.  Figure 4 displays our results.  For the 25-100 AU
semi-major axis regime, we compute an L-dwarf companion fraction upper
limit of 3.3\% (or 4.3\% for a 95\% confidence level).  
     
\subsection[Comparison With Other Published Brown Dwarf Companion Surveys]{Comparison With Other Published Brown Dwarf Companion Surveys}\label{Comparison With Other Published Brown Dwarf Companion Surveys}

Several recently published surveys report brown dwarf companion statistics:  Marcy \& Butler (2000), Gizis et al. (2001), Lowrance et al. (2005), and McCarthy \& Zuckerman (2004).  The population estimate techniques used in these studies vary dramatically in terms of their statistical robustness;  in the following paragraphs we examine the individual survey procedures in order to judge the meaningfulness, and indeed, mathematical rigor, of their results.   

     McCarthy \& Zuckerman (2004), conducting a coronagraphic survey
     for substellar companions to young, northern stars, report a
     highly constrained companion fraction of 1\% $\pm$ 1\% for the 75-300 AU
     semi-major axis region.  However, their conclusions rely on
     several problematic assumptions.  For instance, accurate brown dwarf
     population statistics rely on a rigorous determination of survey
     sensitivity levels.  Such sensitivity estimates in turn rely on
     authors' accurate estimations of target ages.  This is true
     because, for example, a 50 Myr old brown dwarf companion would be much brighter, and therefore easier to detect, than a 10 Gyr companion (see Baraffe et al. [2003] for a quantitative discussion of brown dwarf thermal evolution).  If true target ages turned out to be larger than the authors' estimations, then the derived companion fractions and associated uncertainties would need to be increased.  Metchev (2005) argues that precisely this scenario exists in the McCarthy \& Zuckerman data set.  He points out that McCarthy \& Zuckerman base their age estimates on the assertion that 21\% of solar neighborhood stars possess ages less than 1 Gyr.  This estimate follows the Rocha-Pinto et al. (2000) conclusion that 21\% of stars \textit{observed} in a Rocha-Pinto et al. survey possess ages less than 1 Gyr.  However, Metchev points out that the Rocha-Pinto et al. target sample is flux limited, rather than volume limited.  Hence, the sample biases toward younger, brighter stars.  When Metchev removes this age bias, he derives an updated median target age that is roughly a factor of four greater than the McCarthy \& Zuckerman estimate.

In addition to this error, Metchev also finds potential age errors resulting from the McCarthy \& Zuckerman grouping of visually near-to-each-other target stars into single, co-evolved, co-moving groups.  Metchev argues that, without an independent verification of group membership, or alternatively, an independent age verification, an appreciable fraction of such groups may in fact be older non-members.  Please see Metchev (2005) for a more detailed discussion of this issue.

A final, significant error in the McCarthy \& Zuckerman derivation, an
error which also plagues Lowrance et al. (2005), and Gizis et al. (2001) conclusions, is the incorrect assumption
that all brown dwarf companion orbits are necessarily face-on and
circular.  To illustrate the errors that propagate with such an
assumption, we consider the case of the Gizis et al. (2001) population
estimate for wide-separation brown dwarf companions detected via the
Two Micron All Sky Survey (2MASS; Skrutskie et al. 1997).  Among
its conclusions, the Gizis et al. study reports a 1.4\% $\pm$ 1.1\%
wide-separation (semi-major axis $>$ 1000 AU) L-dwarf companion fraction
to main sequence stars.  We conduct a re-analysis of their 2MASS data
set using Monte Carlo simulations which take into account orbital
projection effects.  Following Gizis et al. (2003) detection
requirements, we assume in our simulations that L-dwarf companions
require a minimum 40-arcsecond separation to be distinguishable from
the parent star flux.  Improving slightly on the Gizis et al. (2003)
angular separation requirements, we also calculate that, even at
$\geq$ 40-arcsecond, there is $\sim$ 3\% chance of a field star
obscuring the companion.  Our re-analysis models then consider
all potential orbits from  100 AU to 40000 AU at log[AU] intervals $\sim$0.1.  

Running our simulations with the afore-mentioned parameters, we
determine that the 1.4\% $\pm$ 1.1\% L-dwarf companion fraction
concluded by Gizis et al.
instead results in a wide-separation (1000 AU - 40000 AU semi-major
axis) L-dwarf companion fraction between 0\% and 4\%.  (A
wide-separation L-dwarf companion fraction of 0\% would correspond to the
cases where published ``wide-separation'' brown dwarf companions
[e.g. Wilson et al. 2001, Scholz et al. 2003] were most likely $<$1000 AU semi-major axis companions
projected onto the $>$1000 AU projected separation space as they
passed through apastron.)  The assertion or refutation of a wide separation brown dwarf desert, based on L-dwarf companion statistics, depends on a few additional factors, such as the relevant comparative stellar fraction, and the extrapolation used to determine the comprehensive brown dwarf population.  But regardless of which of these factors one uses, the fact that our L-dwarf companion fraction includes 0\% as a reasonable possibility means, at the very least, that the Gizis data set cannot support the primary conclusion of the Gizis et al. (2001) paper, ``no brown dwarf desert at wide separations.''

While we refrain from re-deriving statistics for the Lowrance et al. (2005) and
McCarthy \& Zuckerman (2004) data sets, their lack of consideration for
orbital projection effects, and the additional aforementioned age problems for the case of the McCarthy \& Zuckerman paper,  make their results suspect.  Taking this fact under
consideration, we refrain from judging meaningful companion fractions
from these studies until a proper analysis of these systematic errors
has been done.

In contrast to Lowrance et al. (2005),
Gizis et al. (2001), and McCarthy \& Zuckerman (2004), the study by
Marcy \& Butler (2000) includes a significantly more robust account of orbital projection effects.  Being a radial velocity survey, their semi-major axis measurements derive from period, rather than an orbital projection.  Hence, inclination is the dominant orbital bias that they must consider.  Accounting for this bias with analytical approximations, they conclude a brown dwarf companion fraction, for main sequence stars, of $<$1\% for $<$3 AU semi-major axes.  Our companion fraction is consistent with this brown dwarf desert, though we cannot confirm or contradict the extension of this extreme paucity through the 25-100 AU semi-major axis region.

The results from our re-analysis of the Gizis et al. L-dwarf companion
fraction presents a comparison to our intermediate separation survey results.  The 0\%-4\%
companion fraction, derived from that re-analysis, suggests that the
$>$1000 AU L-dwarf companion
fraction may be similar to the intermediate separation (25-100 AU) L-dwarf companion fraction (0.0\%-3.3\%),  derived from our CHAOS survey.

\section{Conclusions}
    Using Monte Carlo simulations to account for orbital projection
    effects, we conclude that observational results from the CHAOS
    survey for brown dwarf (13-73 M$_{Jup}$) companions is most consistent with a brown
    dwarf companion fraction of 0\% - 9.3\%, for the 25-100 AU
    semi-major axis regime.  We compare this value to
    brown dwarf companion fractions reported by Marcy \& Butler (2003)
    for $<$3 AU separations, and find our results to be consistent
    with their value.  However, our uncertainties are too large to either confirm or contradict the continuance of this extreme paucity through the intermediate separation semi-major axis region.  Compared to the published main sequence stellar companion fraction (just above 10\% for 0.08-0.32 M$_{\odot}$ mass companions and a 25-100 AU semi-major axis range [Duquennoy \& Mayor 1991 and Fischer \& Marcy 1992]), our simulations suggest that the brown dwarf companion fraction is smaller than the fraction one would expect by an extrapolation of stellar companion statistics.

For the L-type brown dwarf subset, we
    determine a companion fraction of 0.0\%-3.3\%.  We find that this value is comparable to wide separation (1000 AU - 40000 AU semi-major axis) L-dwarf companion fraction estimates by Gizis et al. (2001), once we correct for systematic observational biases in their study.  As part of this correction re-analysis, we show that the majority of published brown dwarf companion fraction surveys suffer from systematic observational biases (resulting most notably from orbital projection effects) that lead their authors to report companion fraction statistics that are not supported by their observational data.  

\acknowledgements
We thank our anonymous referee for his or her useful suggestions, which improved the manuscript.  J. C. C. and S. S. E. were supported in part by NSF CAREER award AST-0328522.  J. J. S. was supported by the JPL Undergraduate Scholars (JPLUS) program.  The research conducted by J. J. S. and part of the research conducted by J. C. C. were carried out at the Jet Propulsion Laboratory, California Institute of Technology, under a contract with the National Aeronautics and Space Administration.


\begin{figure}
\plotone{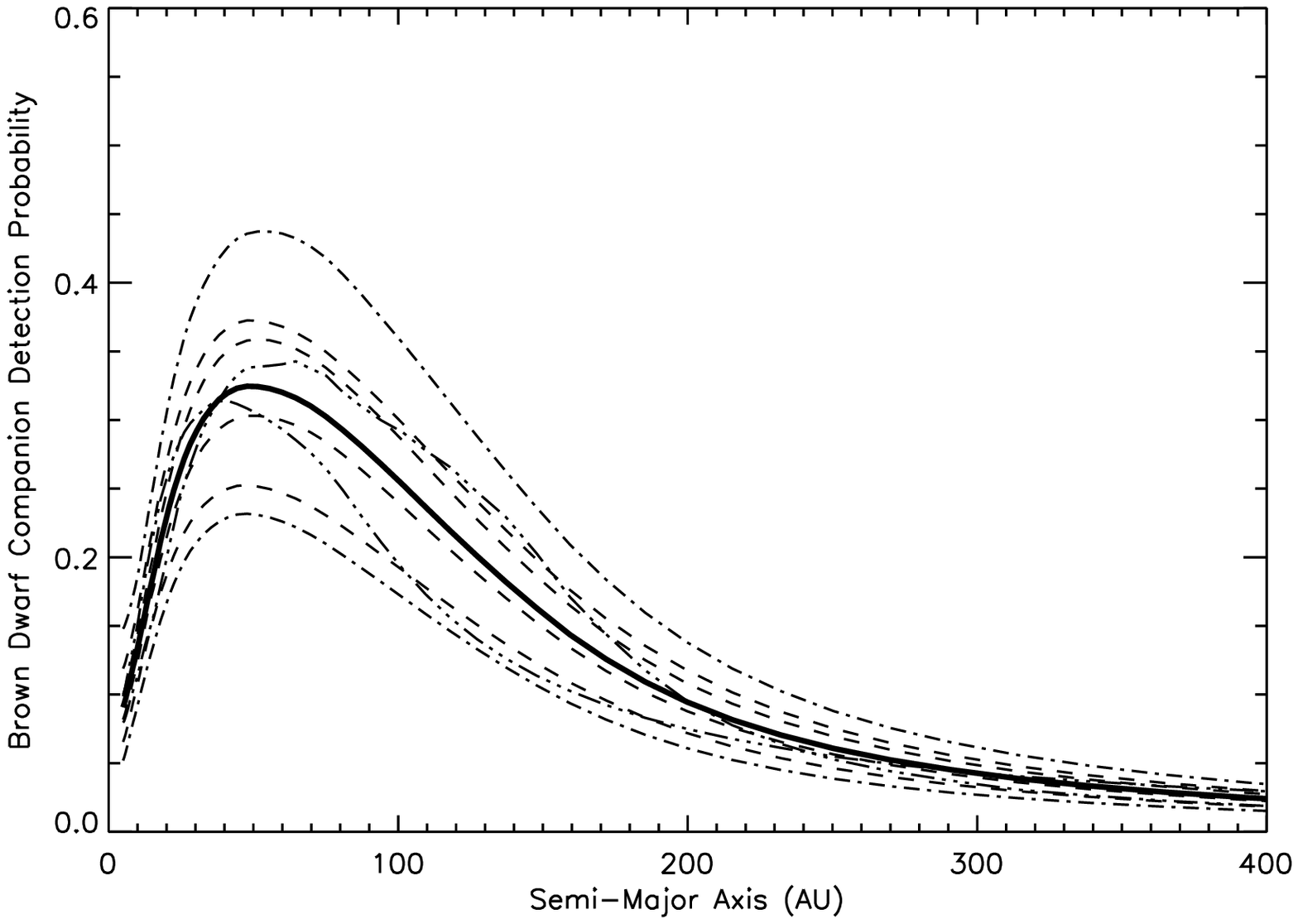}  
\caption{Brown dwarf (13-73 M$_{Jup}$) detection probability versus semi-major axis as
  output from Monte Carlo simulations.  The thick solid curve represents
  output when one assumes a random
  eccentricity, random inclination, random longitude of pericentre, an
  age distribution derived from Table 1, and an initial mass function
  deriving from a median combination of Orion Nebula Cluster (ONC; Slesnick et al. 2004), Taurus (Luhman, 2004) IC 348 (Luhman et al. 2003), Chamaeleon I (Luhman et al. 2005), and Trapezium (Muench et al. 2002) data.
  The \textit{dot-dashed} curves represent the
  \textit{solid} curve data when one skews the median target age by +2
  Gyr (lower curve) and -2 Gyr (upper curve).  The \textit{dashed}
  curves represent detection probabilities using
  Table 1 ages and, from top to bottom, the individual brown dwarf
  mass functions from Trapezium, IC 348, ONC (hidden behind the solid curve), Taurus, and
  Chamaeleon I.  The \textit{dot-dot-dot-dashed} curves represent the \textit{solid}
  curve data, but with eccentricity = 0 (upper curve) and eccentricity
  = 0.9 (lower curve).  All detections
  assume a 5-sigma signal-to-noise ratio.
}
\end{figure}

\begin{figure}
\plotone{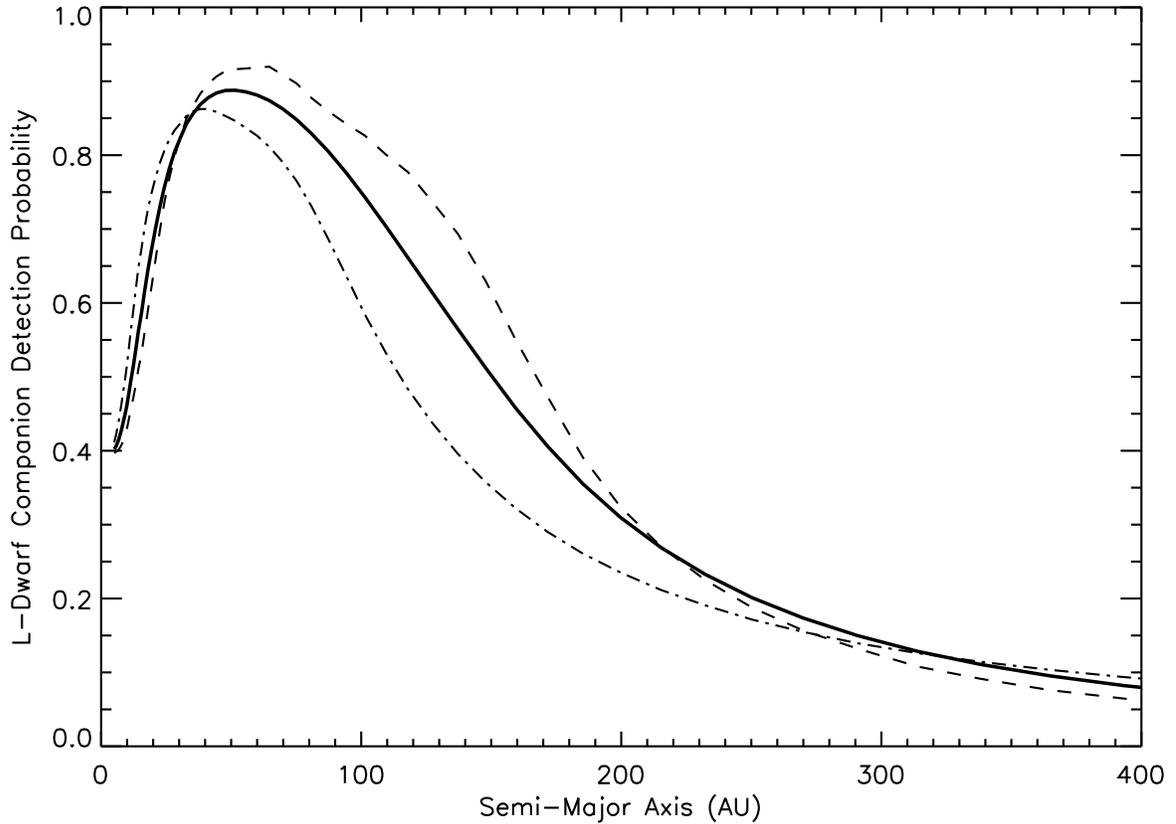}  
\caption{L-dwarf detection probability versus semi-major axis as
  outputted from Monte Carlo simulations.  The solid curve represents
  output when one assumes a random
  eccentricity, random inclination, and random longitude of
  pericentre.  The dashed curves represents the solid curve
  results, but with eccentricity = 0 (upper curve) and eccentricity =
  0.9 (lower curve).  All curves assume an L-dwarf luminosity function
  as given by Cruz et al. (2003).  All detections
  assume a 5-sigma signal-to-noise ratio.
}
\end{figure}

\begin{figure}
\plotone{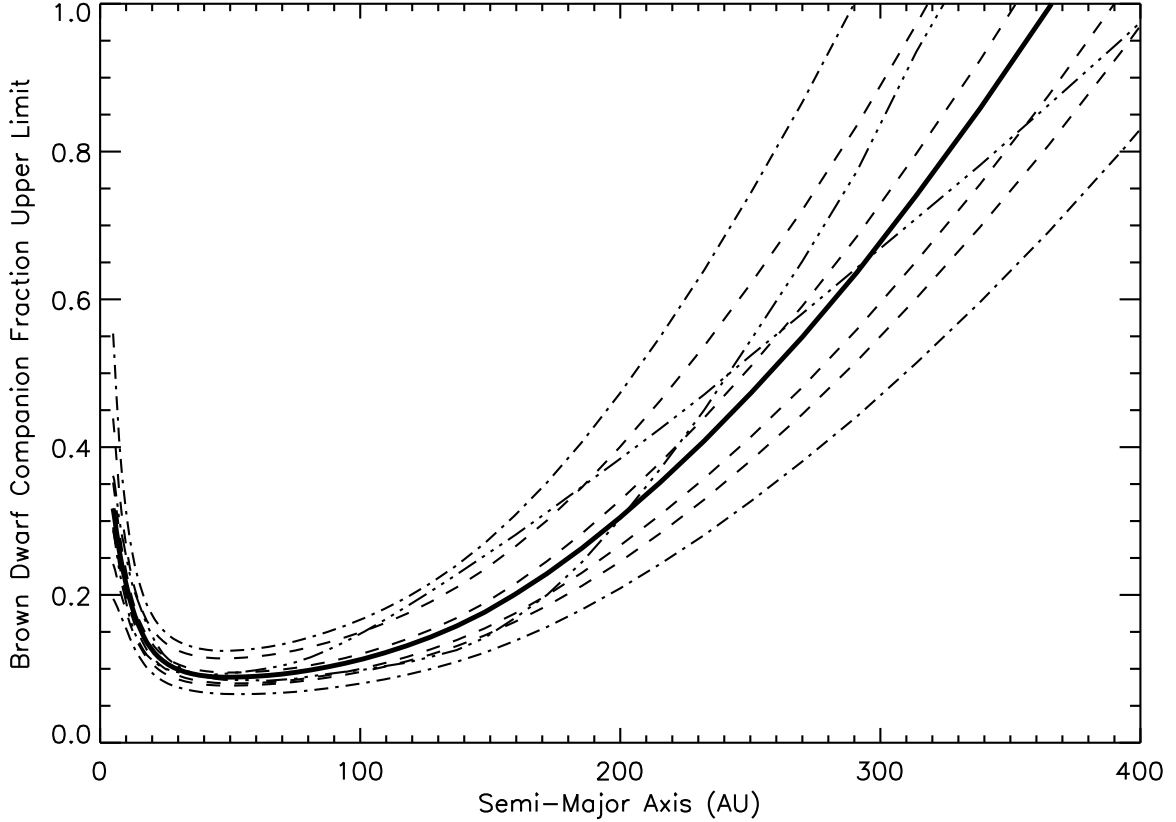}  
\caption{Brown dwarf (13-73 M$_{Jup}$) companion fraction upper limit, at a 90\%
  confidence level, versus semi-major
  axis, as outputted by our Monte Carlo simulations.  The thick solid
  curve represents our preferred parameter set: eccentricity = random;
  inclination = random; longitude of pericentre = random; brown dwarf
  mass function follows a median-combination of ONC (Slesnick et al. 2004), Taurus (Luhman, 2004), IC 348 (Luhman et al. 2003), Chamaeleon I (Luhman et al. 2005), and Trapezium (Muench et al. 2002)
  data; ages derive from Table 1 values.  For this
  preferred case, the plot indicates that at most 9.3\% of target
  stars possess companions, for semi-major axes between 25 and 100 AU.
  The over-plotted curves follow the linestyle convention described in
  Figure 1:  upper \textit{dot-dashed curve} = +2 Gyr offset from Table 1
  derived ages;  lower \textit{ dot-dashed curve} = -2 Gyr offset;
  \textit{dashed} curves represent, from top to bottom, Chamaeleon I, Taurus, ONC
  (hidden behind the solid curve), IC 348, and Trapezium brown dwarf mass
  functions; \textit{dot-dot-dot-dashed} curves represent eccentricity
  = 0 (upper curve as seen from the right) and eccentricity = 0.9
  (lower curve as seen from the right).      
}
\end{figure}

\begin{figure}
\plotone{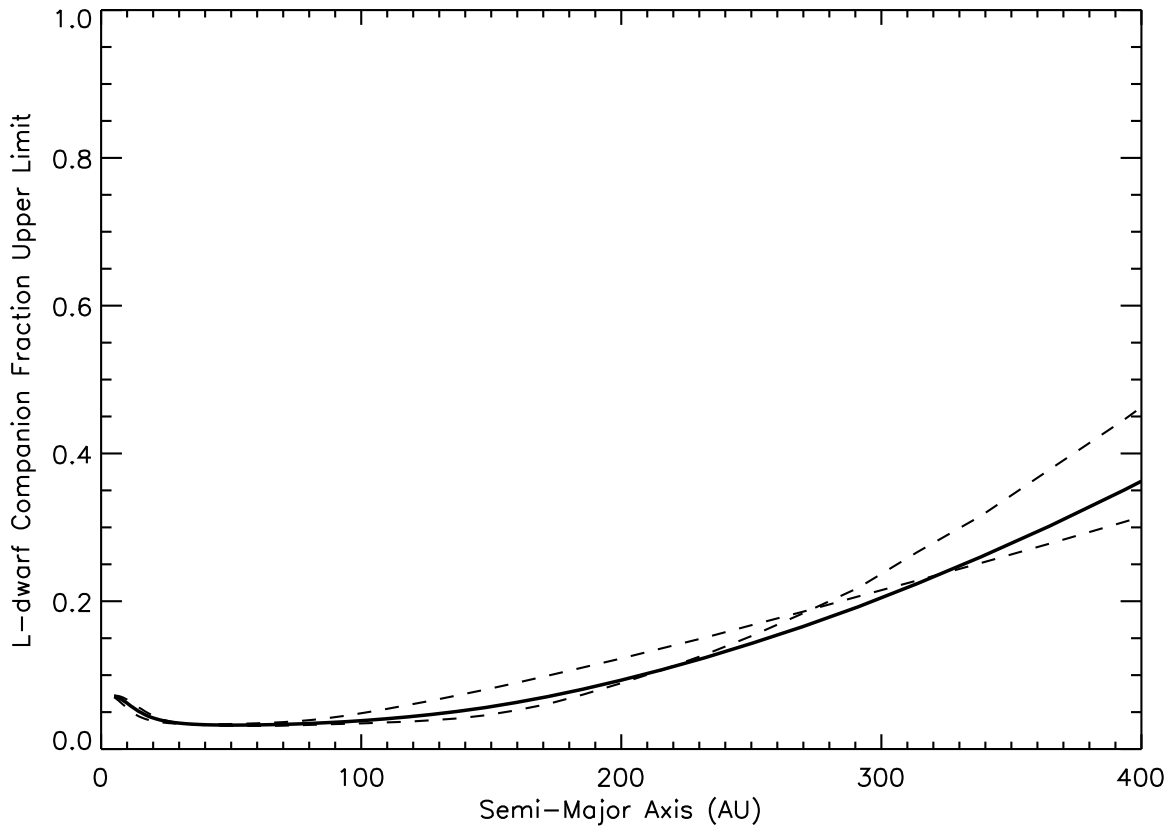}  
\caption{L-dwarf companion fraction upper limit, at a 90\%
  confidence level, versus semi-major
  axis, as outputted by our Monte Carlo simulations.  The thick solid
  curve represents our preferred parameter set: eccentricity = random;
  inclination = random; longitude of pericentre = random.
For this
  preferred case, the plot indicates that at most 3.3\% of target
  stars possess L-dwarf companions, for semi-major axes between 25 and 100 AU.
  The over-plotted dashed curves show the derived companion fraction upper limits when
  eccentricity = 0 (upper curve as seen from the right) and eccentricity = 0.9
  (lower curve as seen from the right).  All curves assume a Cruz et
  al. (2003) L-dwarf luminosity function.  
}
\end{figure}

\begin{deluxetable}{llll}
\tabletypesize{\scriptsize}
\tablecaption{Fe/H Line Strengths}
\tablewidth{0pt}
\tablehead{
\colhead{Star Name} & \colhead{Fe/H} & \colhead{Star Name} & \colhead{Fe/H} \\
}
\startdata
GJ 53 & -0.62 & GJ 211 & -0.20 \\ 
GJ 15 & -0.20 & GJ 166 & -0.19 \\
GJ 27 & -0.32 & GJ 205 & 0.60 \\
GJ 144 & -0.31 & GJ 327 & -0.02 \\
GJ 33 & -0.29 & GJ 475 & 0.02 \\
GJ 411 & -0.20 & GJ 764 & -0.23 \\
GJ 434 & -0.40 & GJ 388 & -0.22 \\
GJ 631 & 0.01 & GJ 368 & -0.04 \\
GJ 488 & 0.10 & GJ 395 & -0.23 \\
GJ 75 & 0.36 & GJ 451 & -1.50 \\
GJ 222 & 0.25 & GJ 534 & 0.44 \\
GJ 183 & 0.02 & GJ 502 & 0.19 \\
GJ 212 & -0.20 & GJ 549 & -0.05 \\
GJ 848 & -0.10 & GJ 449 & 0.33 \\
GJ 92 & -0.43 & GJ 603 & -0.40 \\
GJ 892 & 0.00 & GJ 678 & 0.02 \\
GJ 673 & 0.40 & GJ 807 & 0.13 \\
GJ 41 & 0.10 & GJ 706 & -0.30 \\
GJ 324 & -0.15 & GJ 695 & 0.10 \\
GJ 380 & 0.28 & GJ 598 & -0.04 \\
\enddata
\end{deluxetable}


\begin{thebibliography}{}

\bibitem[Baraffe et al. (2003)]{bet03}Baraffe, I., Chabrier, G.,
  Barman, T. S., Allard, F., \& Hauschildt, P. H. 2003, A\&A, 402, 701
\bibitem[Bate et al. (2002)]{bet02} Bate, M. R., Bonnell, I. A., \&
  Bromm, V. 2002, \mnras, 332, 65
\bibitem[Boss et al. (2001)]{bet01} Boss, A. P. 2001, \apj, 551, 167
\bibitem[Burgasser (2004)]{b04} Burgasser, A. J. 2004, \apjs, 155, 191

\bibitem[Carson (2005)]{c05}  Carson, J. C. 2005, Ph.D. thesis,
  Cornell Univ
\bibitem[Carson et al. (2005)]{c05b}  Carson, J. C., Eikenberry, S. S.,
  Brandl, B. R., Wilson, J. C., \& Hayward, T. L. 2005, \aj, 130, 1212 
\bibitem[Cayrel de Strobel et al. (1997)]{cds97}Cayrel de Strobel, G., Soubiran, C., Friel, E. D., Ralite, N., \& Francois, P., 1997, A\&AS, 124, 299
\bibitem[Cayrel de Strobel et al. (2001)]{cet01}Cayrel de Strobel, G., Soubiran, C., \& Ralite, N., 2001, A\&A, 373, 159
\bibitem[Clarke et al. (2004)]{cet04}
    Clarke, C.,  Reipurth, B., \& Delgado-Donate, E. 2004, RevMexAA,
    21, 184
\bibitem[Cruz et al. (2003)]{cet03}Cruz, K. L., Reid, I. N., Liebert,
  J., Kirkpatrick, J. D., \& Lowrance, P. J. 2003, \aj, 126, 2421

\bibitem[Duquennoy \& Mayor (1991)]{dm91}Duquennoy, A. \& Mayor, M. 1991, A\&A, 248, 485

\bibitem[Fischer \& Marcy (1992)]{fm92}Fischer, D. A. \& Marcy, G. W. 1992, \apj, 396, 178

\bibitem[Gizis et al. (2001)]{get01}Gizis, J. E., Kirkpatrick, J. D.,
  Burgasser, A., Reid, I. N., Monet, D. G., Liebert, J.,
  \& Wilson, J. C. 2001, \apj, 551, 163

\bibitem[Hayward et al.(2001)]{het01}Hayward, T. L., Brandl, B.,
  Pirger, B., Blacken, C., Gull, G. E., Schoenwald, J., \& Houck,
  J. 2001, \pasp, 113, 105

\bibitem[Lowrance et al. (2005)]{lowet05} Lowrance, P. J., et al. 2005, \aj, 130, 1845
\bibitem[Luhman (2004)]{l04} Luhman, K. L. 2004, \apj, 617, 1216
\bibitem[Luhman et al. (2003)]{let03} Luhman, K. L., Stauffer, J. R., Muench, A. A., Rieke, G. H., Lada, E. A., Bouvier, J., \& Lada, C. J. 2003, \apj, 593, 1093
\bibitem[Luhman et al. (2005)]{let05} Luhman, K. L., Fazio, G., Megeath, T., Hartmann, L., \& Calvet, N.  2005, Mem. S.A.IT., 76, 285

\bibitem[Marcy \& Butler(2000)]{mar00}Marcy,~G.~W. \& Butler,~R.~P.
    2000, \pasp, 112, 137
\bibitem[McCarthy \& Zuckerman (2004)]{mz04}McCarthy,~C. \&
    Zuckerman,~B. 2004, \aj, 127, 2871
\bibitem[Metchev (2005)]{m05} Metchev, S. A. 2005, Ph.D. thesis, California Institute of Technology
\bibitem[Muench et al. (2002)]{met02} Muench, A. A., Lada, E. A., Lada, C. J., \& Alves, J. 2002, \apj, 573, 366

\bibitem[Reipurth \& Clarke (2001)]{rc01}Reipurth, B. \& Clarke,
  C. 2001, \aj, 122, 432 
\bibitem[Rocha-Pinto et al. (2000)]{ret00}Rocha-Pinto, H. J., Maciel, W. J., Scalo, J., \& Flynn, C. 2000, A\&A, 358, 850 

\bibitem[Skrutskie et al. (1997)]{set97}Skrutskie, M. F., et al. 1997,
  in The Impact of Large-Scale Near-IR Sky Surveys, ed. F. Garzon
  (Dordrecht: Kluwer), 25
\bibitem[Slesnick et al. (2004)]{set04} Slesnick, C. L., Hillenbrand, L. A., \& Carpenter, J. M. 2004, \apj, 610, 1045
\bibitem[Song et al. (2004)]{songetal04} Song, I., Zuckerman, B., \& Bessell, M. S. 2004, \apj, 614, 125

\bibitem[Tokovinin (1992)]{t92}Tokovinin, A. A. 1992, A\&A, 256, 121

\bibitem[Troy et al.(2000)]{tet00}Troy et al. 2000, SPIE, 4007, 31

\bibitem[Wilson et al. (2001)]{wet01}Wilson, J. C., Kirkpatrick,
  J. D., Gizis, J. E., Skrutskie, M. F., Monet, D. G., \& Houck,
  J. R. 2001, \aj, 122, 1989

\end{thebibliography}
\end{document}